\documentclass{article}
\usepackage{pdfpages}
\usepackage{xcolor}
\usepackage[mathlines]{lineno}
\usepackage{graphicx}
\usepackage{cite}
\usepackage{caption}
 
\usepackage{subfig} 
\usepackage{subcaption}
\usepackage{authblk}
\usepackage{amsmath}
\usepackage[top=5cm, bottom=5cm, left=3cm, right=2cm, headsep=1.5cm, heightrounded=true]{geometry}
\usepackage{fancyhdr}
\graphicspath{ {{images/}} }

\title{Dark Matter Signatures in Black Hole Thermodynamics and Information Recovery}

\author[1,2]{Yahya Ladghami\thanks{\texttt{yahya.ladghami@ump.ac.ma}}}
\author[1,2]{Brahime Asfour\thanks{\texttt{brahim.asfour@ump.ac.ma}}}
\author[1,2]{Taoufik Ouali\thanks{\texttt{t.ouali@ump.ac.ma}}}

\affil[1]{Laboratory of Physics of Matter and Radiation, Mohammed I University, BP 717, Oujda, Morocco}
\affil[2]{Astrophysical and Cosmological Center, BP 717, Oujda, Morocco}

\begin{document}
	\maketitle
	\begin{abstract}
	In this paper, we investigate the thermodynamic properties and information recovery of Schwarzschild and Reissner–Nordström black holes surrounded by perfect fluid dark matter. We show that, while the Bekenstein–Hawking entropy remains unchanged, dark matter significantly modifies the Hawking temperature, introducing a positive contribution that enhances thermal effects, particularly for small black holes. We find that the phase structure is preserved, where Schwarzschild black holes remain unstable and Reissner–Nordström black holes exhibit a standard small/large black hole transition, in which small black holes are stable and large black holes are unstable. Furthermore, we demonstrate that dark matter accelerates Hawking evaporation, reducing black hole lifetimes.
	We further investigate the black hole information loss paradox using the island formula. In the absence of islands, the entanglement entropy of Hawking radiation grows linearly with time and diverges at late times, violating unitarity. By including island contributions, the entanglement entropy of Hawking radiation saturates at twice the Bekenstein–Hawking entropy, reproducing the Page curve and restoring information recovery for both Schwarzschild and Reissner–Nordström black holes surrounded by perfect fluid dark matter. We derive analytical expressions for the Page time and demonstrate that it is directly determined by the thermodynamic parameters of the black hole. Furthermore, we establish a correspondence between thermodynamics and information recovery by showing that the Page time is controlled by the Hawking temperature and the event horizon. Furthermore, we found that the presence of dark matter reduces the Page time, thereby accelerating the recovery of information.
	
	\end{abstract}
	
\section{Introduction}
In recent years, two very important achievements have been made in the study of the Universe: the detection of gravitational waves by the LIGO/Virgo collaborations \cite{I1}, where they were emanating from the merger of two black holes, and the capture of the shadow images of M87 and Sagittarius A* black holes by the Event Horizon Telescope \cite{I2, I3}. These discoveries supported the development of black hole physics. But, before these discoveries, black holes already had a very important role in understanding our Universe, because black holes act as natural laboratories for studying astrophysical phenomena and as theoretical laboratories for testing new ideas about quantum gravity and the unification of forces. Specifically, with the discovery by Hawking of black hole evaporation through the emission of blackbody radiation, called Hawking radiation \cite{Hawking:1975vcx}. After this discovery, our view of black holes changed. Indeed, black holes are not only strong gravitational systems, but also thermodynamic systems characterized by a temperature and an entropy \cite{22A}, and they obey laws that closely resemble those of classical thermodynamics \cite{3a,3aa}. This thermodynamic perspective has uncovered a remarkably rich structure, especially for black holes in asymptotically Anti-de Sitter (AdS) spacetime, including the Hawking-Page transition \cite{Hawking:1982dh,Eune:2013qs,Li:2020khm}, critical behavior \cite{a1,a2,a3,A0,A1,A11,A2,A4,A44,A5,L,He,Di,3b,3BB}, and multicritical points \cite{W,Wu}.
\\

Despite these developments, Hawking radiation gives rise to one of the most profound problems in modern theoretical physics: the black hole information loss paradox. This paradox originates from a conflict between black hole evaporation and the principles of quantum mechanics \cite{16a}. Hawking radiation is thermal and appears to carry no information about the matter that formed or entered the black hole. Consequently, if a black hole evaporates completely, the initial pure quantum state would evolve into a mixed state, violating the principle of unitarity \cite{Mathur:2009hf}. This  loss of information is known as the black hole information loss paradox.
 \\

 To address this issue, Page proposed that the entanglement entropy of Hawking radiation should follow a specific time evolution consistent with unitary quantum mechanics \cite{17aa}. According to this scenario, the entropy initially increases, reaches a maximum value at the Page time, and subsequently decreases as information begins to emerge in the Hawking radiation. This characteristic behavior is known as the Page curve. More recently, a promising resolution of the information loss paradox has been proposed through the island formula \cite{18aa}. In this framework, the contribution of a regions inside the black hole, called islands, is included in the computation of the entanglement entropy of Hawking radiation. Remarkably, this approach successfully reproduces the Page curve and restores unitary evolution. The validity of the island formula has been demonstrated for a wide variety of black hole solutions, including Schwarzschild \cite{19,19a}, Reissner-Nordström \cite{20}, BTZ \cite{ytr}, and black holes in conformal Killing gravity \cite{yll}.
 \\

On the other hand, dark matter constitutes one of the most significant mysteries in contemporary physics. While its existence is strongly supported by astrophysical and cosmological observations, its fundamental nature remains unknown. Dark matter is considered as a fundamental component of our Universe, constituting approximately 26–27\% of its total mass-energy content according to the standard model of cosmology \cite{Planck:2018vyg}. In realistic astrophysical environments, black holes are not isolated systems but are instead embedded in matter distributions, including dark matter halos. For this reason, understanding how such an environment influences the geometry and thermodynamics of black holes is very important for establishing a more complete and physically relevant description of black holes. The modeling of dark matter as a perfect fluid is a particularly useful approach \cite{Rahaman:2010xs}, which consists of describing dark matter as a perfect fluid surrounding  black holes. Within this framework,  Einstein field equations admit solutions in which the presence of dark matter modifies the metric function through a logarithmic correction term \cite{Xu:2016ylr}. This modification leads to changes in the near-horizon structure of spacetime. Consequently, all thermodynamic quantities derived from the horizon geometry, such as temperature, heat capacity, and free energy, are expected to be affected, where dark matter leaves its signatures on the thermodynamics and evaporation of black holes. Furthermore, this field of study, namely the interaction between black holes and dark matter, is very rich, with many works having been devoted to it, such as the dynamics of test particles around a Bardeen black hole surrounded by dark matter \cite{Feng:2024tgc}, the influence of perfect fluid dark matter on the shadow observables of black holes \cite{DMS1,DMS2,DMS3,DMS4,DMS5}, the thermodynamic topological classification of black holes surrounded by perfect fluid dark matter \cite{TDM,TDM1}, and the extended phase space thermodynamics of black holes surrounded by perfect fluid dark matter \cite{EPSTDM,EPSTDM1,EPSTDM3}.
\\

Motivated by these developments,  we investigate the signatures of dark matter on both black hole thermodynamics and information recovery. Specifically, we study the thermodynamic properties of Schwarzschild and Reissner-Nordström black holes surrounded by perfect fluid dark matter and analyze information recovery through the island formula and the Page time. Furthermore, we explore the correspondence between thermodynamics and information recovery by examining how thermodynamic behavior influences the Page time and the dynamics of information retrieval, as well as how dark matter affects these processes.
\\

The paper is organized as follows. In Sec.~\ref{sec2}, we review the geometry of black holes surrounded by perfect fluid dark matter. In Sec.~\ref{sec3}, we analyze the thermodynamics and stability of Schwarzschild black holes. In Sec.~\ref{sec4}, we extend the discussion to Reissner-Nordström black holes and investigate their phase structure. In Sec.~\ref{sa}, we study information recovery through the island formula. In Sec.~\ref{saa}, we explore the correspondence between thermodynamics and information recovery for black holes surrounded by perfect fluid dark matter. Finally, in Sec.~\ref{sec5}, we present our conclusions and discuss possible future directions. Throughout this work, we adopt natural units in which $G=\hbar=c=k_{\mathrm{B}}=1$.

\section{Black Holes Surrounded by Perfect Fluid Dark Matter}
\label{sec2}

In this section, we review the solutions for  Schwarzschild and Reissner-Nordström black holes surrounded by a perfect fluid dark matter. The metric function \(f(r)\) for charged black holes surrounded by dark matter is derived by solving the Einstein field equations
\begin{equation}
	R_{\mu\nu} - \frac{1}{2} g_{\mu\nu} R = 8 \pi T_{\mu\nu},
\end{equation}
where the energy-momentum tensor, $T_{\mu\nu}$, contains the contributions of the Maxwell field and a perfect dark matter fluid.   For a static and spherically symmetric spacetime, described by the metric
\begin{equation}
	ds^2 = -f(r) dt^2 + \frac{1}{f(r)} dr^2 + r^2 (d\theta^2 + \sin^2\theta \, d\phi^2).
\end{equation}
The contribution from the Maxwell field to the total energy-momentum tensor, \(T_{\mu\nu}\), is described by
\begin{equation}
	T^{\text{Maxwell}}_{\mu\nu} = \frac{1}{4\pi} \left( F_{\mu\alpha} F^\alpha_\nu - \frac{1}{4} g_{\mu\nu} F_{\alpha\beta} F^{\alpha\beta} \right),
\end{equation}
with $F_{\mu\nu}$ being the electromagnetic field strength tensor. For a spherically symmetric electric field, the nonzero components are \(F_{tr}\), leading to \cite{Gao:2023ltr}
\begin{equation}
	T_t^{t\, \text{(Maxwell)}} = T_r^{r\, \text{(Maxwell)}} = -\frac{Q^2}{8\pi r^4}, \quad \text{and} \quad 
	T_\theta^{\theta\, \text{(Maxwell)}} = T_\phi^{\phi\, \text{(Maxwell)}} = \frac{Q^2}{8\pi r^4}.
\end{equation}
Here \(t\) and \((r,\theta,\phi)\) denote the time and spherical coordinates, respectively, and \(Q\) is the electric charge. The dark matter contribution to the total energy-momentum tensor is given by \cite{Das:2020yxw,Gao:2023ltr}
\begin{equation}
	T_{\mu\nu}^{\text{DM}} = \text{diag}(-\rho, P_r, P_\theta, P_\phi),
\end{equation}
where the energy density \(\rho\) and the pressures \(P_r\), \(P_\theta\), and \(P_\phi\) are given by 
\begin{equation}
	\label{P}
	\rho = -P_r = \frac{\lambda}{8\pi r^3}, \qquad \text{and} \qquad 
	P_\phi = P_\theta = \frac{\lambda}{16\pi r^3}.
\end{equation}

Substituting the expression for \(T_{\mu\nu}\) into the Einstein equations, the nonzero components yield coupled differential equations for \(f(r)\). For the \(R^t_t\) and \(R^r_r\) components, we obtain
\begin{equation}
	\frac{f'}{r} + \frac{f - 1}{r^2} = -\frac{Q^2}{r^4} + \frac{\lambda}{r^3},
\end{equation}
where the prime denotes the derivative with respect to $r$. while for \(R^\theta_\theta\) and \(R^\phi_\phi\), the equation becomes
\begin{equation}
	\frac{f''}{2} + \frac{f'}{r} = \frac{Q^2}{r^4} + \frac{\lambda}{2r^3}.
\end{equation}
Solving these equations, the metric function is obtained as  \cite{Das:2020yxw,Gao:2023ltr}
\begin{equation}
	\label{Ei}
	f(r) = 1 - \frac{2M}{r} + \frac{Q^2}{r^2} + \frac{\lambda}{r} \ln\left(\frac{r}{\lambda}\right),
\end{equation}
where \(M\) is the black hole mass,  and \(\lambda\) characterizes the density of the surrounding dark matter. This result highlights the interplay between black holes mass, electric charge, and dark matter effects, with the logarithmic term specifically reflecting the influence of dark matter on the spacetime geometry of the black hole.

In this study, we focus on the thermodynamics of  Schwarzschild and RN black holes surrounded by dark matter. The Schwarzschild solution is recovered by setting \(Q=0\) in Eq.~\eqref{Ei}.



	\section{ Schwarzschild Black Holes}
	\label{sec3}
In this section, we examine the role of dark matter in shaping the thermodynamic behavior, stability, and phase structure of Schwarzschild black holes.  We can find the black hole mass from the metric function Eq.~\eqref{Ei} with $Q=0$, by solving $f(r_0)=0$, where $r_0$ is the event horizon of the black hole. We find the following expression  
\begin{equation}
	\label{Mass}
	M = \frac{r_0}{2} + \frac{\lambda}{2} \log \left( \frac{r_0}{\lambda} \right).
\end{equation}

We can derive the Hawking temperature as follows  
\begin{equation}
	\label{Ta} 
	T= \frac{f'(r_0)}{4\pi}= \frac{\lambda + r_0}{4 \pi r_0^2}.
\end{equation}
 Furthermore, we can calculate the black hole entropy as follows  
\begin{equation}
	\label{BHE}
	S= \int_{0}^{r_0} \frac{dM}{T}= \pi r_0^2.
\end{equation}  
This entropy recovers the Bekenstein-Hawking entropy law, i.e. the  dark matter surrounding black holes does not affect this law. To study the effect of this dark matter on the thermodynamic properties of the black hole, we treat the dark matter density as a thermodynamic variable and define its conjugate quantity, $\Psi$. We write the first law of thermodynamics as follows

\begin{equation}
	dM= TdS + \Psi d\lambda.
\end{equation}
From this first law,  the expression of the dark matter potential writes  
\begin{equation}
	\Psi= \left(\frac{\partial M}{\partial \lambda} \right)_S = \frac{1}{2} \log \left( \frac{r_0}{\lambda} \right) - \frac{1}{2}.
\end{equation}  
The Smarr relation reads consequently through these thermodynamic quantities, as
\begin{equation}
	M = 2 T S + \Psi \lambda.
\end{equation}

\subsection*{Phase Structure}

The phase structures of Schwarzschild black holes are studied by means of their thermal evolutions and stabilities. To study the stability, we consider the heat capacity which is given by  
\begin{equation}
	C_\lambda= T \left( \frac{\partial S}{\partial T}\right)_\lambda= -\frac{2\pi r_0^2 (r_0 + \lambda)}{r_0 + 2\lambda}.
\end{equation}  
  
From the expression of the heat capacity, we observe that this black hole exhibits only one thermodynamic phase, which is unstable because the heat capacity is negative. Therefore, dark matter does not affect the stability or the phase structure of Schwarzschild black holes.
\\

\begin{figure}[htp]
	\centering
	\includegraphics[width=0.7\linewidth]{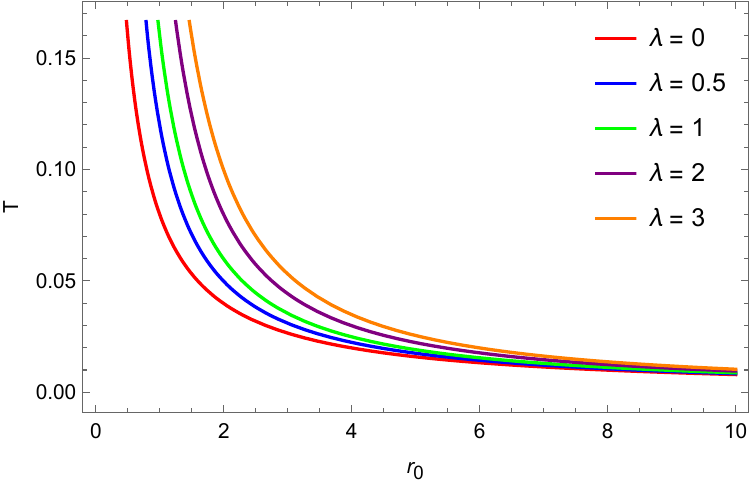}
	\caption{Thermal evolution of Schwarzschild black holes in terms of the event horizon for different values of the dark matter parameter.}
	\label{fig:trs}
\end{figure}
Fig.~\ref{fig:trs} represents the thermal evolution of  black holes for different values of the dark matter parameter. We observe only one thermodynamic phase, where the Hawking temperature decreases with increasing event horizon radius. Consequently, small black holes have a higher temperature than large black holes.
\\

Regarding the impact of dark matter, represented by the parameter $\lambda$, we observe that dark matter affects the value of the Hawking temperature. Indeed, the temperature increases as the parameter $\lambda$ increases. Since this parameter is related to the dark matter density, we conclude that the dark matter density influences the Hawking temperature. This effect is more pronounced for small black holes and becomes weaker for large black holes. Therefore, a higher dark matter density corresponds to a higher Hawking temperature, particularly for small black holes.

\subsection*{Lifetime of Black Holes}

Black holes undergo evaporation through the emission of Hawking radiation, ultimately leading to their complete disappearance. The duration of this process, commonly referred to as the black hole lifetime, depends on the initial mass of the black hole and the associated Hawking temperature. The evaporation process can be modeled as blackbody radiation as follows \cite{ASMAA}

\begin{equation}
	\label{tls}
	\frac{d M}{d t} = - \sigma A T^4,
\end{equation}
where $\sigma$ represents the Stefan-Boltzmann constant and $A$ is the area of the black hole.
\\

On the other hand, we found in the previous part of this section that dark matter affects the Hawking temperature of black holes, where an increase in the dark matter density leads to an increase in the Hawking temperature. These results lead us to ask how the lifetime of black holes is affected by the dark matter density. To determine this effect, we calculate the lifetime of black holes, $t_{evp}$, surrounded by dark matter. Using Eq.~\eqref{tls}, we obtain
\begin{equation}
	\label{eqtevp}
	\begin{aligned}
		t_{evp}
		&=\int_{M_0}^{0} -\frac{dM}{\sigma A T^4}  \\
		&= \frac{16}{3\,\sigma}\,\pi^3
		\left[
		\frac{r_0 \left(2 r_0^4 - 5 r_0^3 \lambda + 20 r_0^2 \lambda^2 + 90 r_0 \lambda^3 + 60 \lambda^4 \right)}
		{(r_0 + \lambda)^2}
		+ 60 \lambda^3 \left(\ln \lambda - \ln (r_0 + \lambda)\right)
		\right],
	\end{aligned}
\end{equation}
where $M_0$ and $r_0$ are the initial mass and event horizon radius of the black hole, respectively.\\

\begin{figure}[htp]
	\centering
	\includegraphics[width=0.7\linewidth]{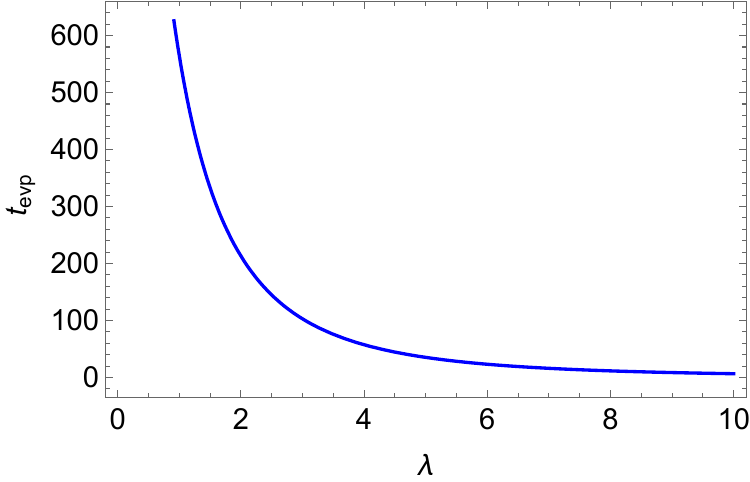}
	\caption{Lifetime of black holes surrounded by dark matter as a function of the dark matter parameter, for $r_0=2$ and $\sigma=1$.}
	\label{fig:tevp}
\end{figure}
 Fig.~\ref{fig:tevp} represents the evolution of the lifetime of a black hole surrounded by dark matter as a function of the dark matter parameter for a fixed event horizon $r_0=2$ and Stefan-Boltzmann constant $\sigma=1$. This figure shows how dark matter affects black hole evaporation and its lifetime. We observe that the lifetime decreases as the dark matter parameter increases. This means that the presence of dark matter around black holes enhances the evaporation process. Therefore, black holes surrounded by dark matter have a shorter lifetime compared to black holes without dark matter. 

This phenomenon, in addition to the effect of dark matter on the Hawking temperature, shows how the environment influences the thermodynamic behavior of black holes. We conclude that black hole evaporation and thermal evolution are not only determined by the intrinsic properties of the black hole, but can also be affected by the surrounding environment.

\section{ Reissner-Nordström Black Holes}
\label{sec4}

In this section, we extend our framework to charged black holes and examine how dark matter modifies the thermodynamic behavior and phase structure of Reissner-Nordström black holes. We first derive the mass expression from Eq.~\eqref{Ei}, which takes the form
\begin{equation}
	M_{RN}= \frac{r_+}{2} + \frac{Q^2}{2 r_+} + \frac{\lambda}{2}\ln\left(\frac{r_+}{\lambda}\right),
\end{equation}
where $r_+$ is the event horizon of the RN black hole surrounded by dark matter. 

To study the impact of dark matter on black hole thermodynamics, we need the Hawking temperature, which is given by
\begin{equation}
	\label{TRNz}
	T_{RN}= \frac{f'(r_+)}{4\pi}= \frac{1}{4 \pi r_+} 
	-\frac{Q^2}{4 \pi r_+^3} +  \frac{ \lambda}{4 \pi r_+^2}.
\end{equation}

From this expression, we observe that the Hawking temperature of RN black holes is also affected by the dark matter environment. When $\lambda \to 0$, we recover the Hawking temperature of the RN black hole without dark matter. Additionally, from this expression, we observe that dark matter has a positive contribution to the Hawking temperature. This means that the presence of dark matter around RN black holes increases their temperature and enhances their evaporation, similar to Schwarzschild black holes. 

Additionally, there exists an extremal limit of RN black holes, where the black hole does not emit any radiation, corresponding to zero Hawking temperature. This case corresponds to the extremal event horizon, $r_{ext}$, which can be expressed as follows
\begin{equation}
	r_{ext}=\frac{1}{2}\left(-\lambda + \sqrt{4 Q^2 + \lambda^2}\right).
\end{equation}
Like Schwarzschild black holes surrounded by dark matter, the entropy of RN black holes does not violate the Bekenstein--Hawking area law of entropy. Indeed, the entropy of the RN black hole is given by
\begin{equation}
	S= \int_{0}^{r_+} \frac{d M_{RN}}{T_{RN}}= \pi r_+^2,
\end{equation}
 Therefore, we conclude that the dark matter contribution does not change the expression of black hole entropy. \\

The thermodynamic first law of RN black holes surrounded by dark matter is given by
\begin{equation}
	dM_{RN}= T_{RN} dS + \Phi dQ+\Psi d\lambda,
\end{equation}
where  $\Phi$ is the electric potential of black hole.
Thus, we can express the electric potential, $\Phi$, and the dark matter potential, $\Psi$, as follows
\begin{equation}
	\Phi= \left(\frac{\partial M_{RN}}{\partial Q} \right)_{S,\lambda} = \frac{Q}{r_+},
\end{equation} 
and
\begin{equation}
	\Psi=\left(\frac{\partial M_{RN}}{\partial \lambda} \right)_{S,Q} = \frac{1}{2} \log \left( \frac{r_+}{\lambda} \right) - \frac{1}{2}.
\end{equation}

\subsection*{Phase Structure}

Now, we study the phase structure and critical phenomena of RN black holes, and we investigate the role of dark matter in the thermodynamic behavior of RN black holes. For this, we find the quantities corresponding to the critical phenomena by solving the following equations  
\begin{equation}
	\left(\dfrac{\partial T_{RN}}{\partial r_+} \right)_{Q, \lambda}= 	\left(\dfrac{\partial^2 T_{RN}}{\partial r_+^2} \right)_{Q, \lambda}=0.
\end{equation}
These equations do not have any real and positive solutions, i.e. there are no critical phenomena or second-order phase transitions for RN black holes surrounded by dark matter. This means that there does not exist any critical phenomena or second-order phase transition for RN black holes surrounded by dark matter. 

To investigate the first-order phase transition, we solve the following expression  
\begin{equation}
	\left(\dfrac{\partial T_{RN}}{\partial r_+} \right)_{Q, \lambda}=0,
\end{equation}
This equation determines the event horizon radius, $r_t$, corresponding to the first phase transition and can be expressed as 
\begin{equation}
	\label{rpt}
	r_t= -\lambda + \sqrt{3 Q^2 + \lambda^2}.
\end{equation}

At this  event horizon radius, a phase transition occurs between small and large black holes, corresponding to $r_{\mathrm{ext}} < r_+ < r_t$ and $r_+ > r_t$, respectively. Through Eq.~\eqref{rpt}, we observe that the phase transition phenomenon is directly related to the electric charge and the dark matter parameter. This shows how  black hole characteristics, such as charge, and the environment around the black hole affect the thermodynamic behavior.   
\\

\begin{figure}[htp]
	\centering
	\includegraphics[width=0.4\linewidth]{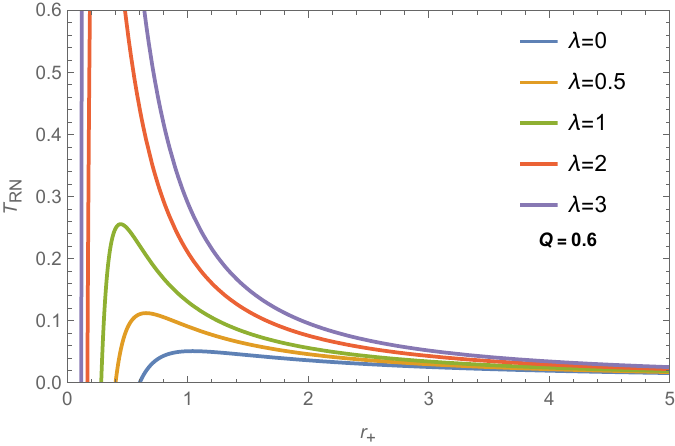}
		\includegraphics[width=0.4\linewidth]{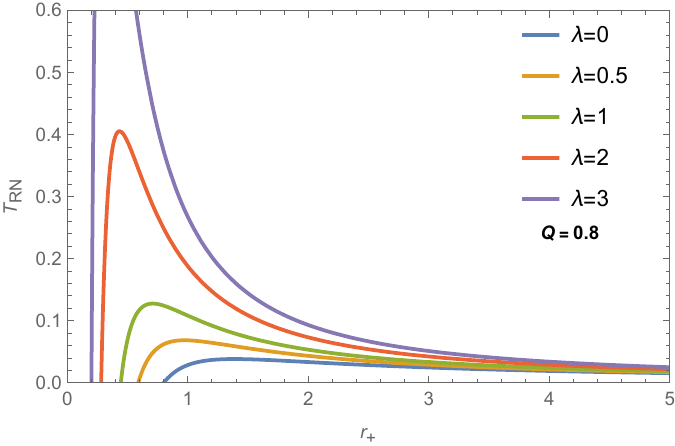}
			\includegraphics[width=0.4\linewidth]{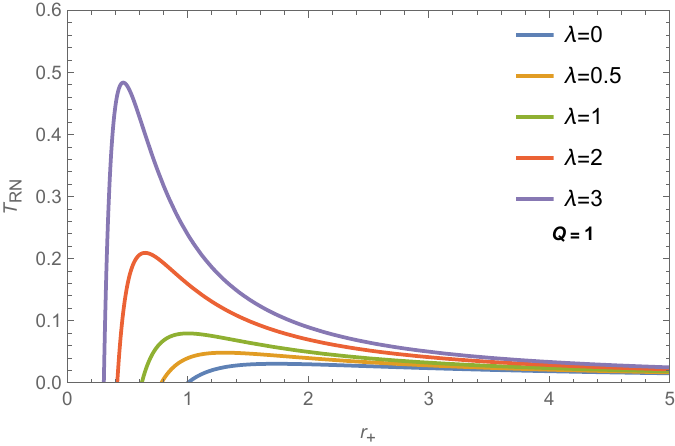}
				\includegraphics[width=0.4\linewidth]{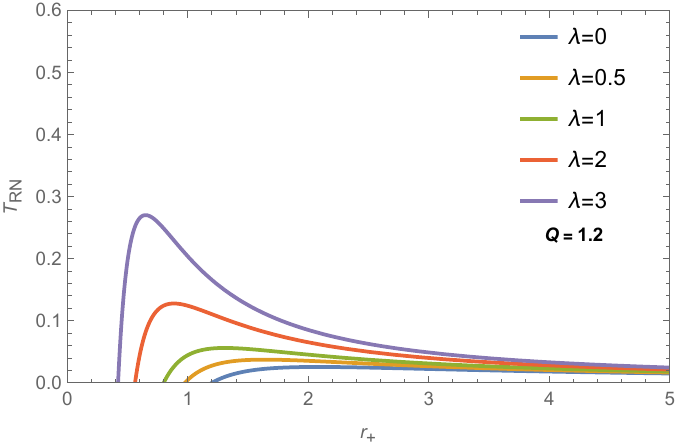}
	\caption{Hawking temperature of RN black holes surrounded by dark matter for different values of $\lambda$ and $Q$.}
	\label{fig:trn}
\end{figure}

Fig.~\ref{fig:trn} shows the Hawking temperature $T_{RN}$ of the Reissner-Nordstr\"om black hole surrounded by dark matter as a function of the event horizon radius $r_+$, for several values of the dark matter parameter $\lambda$ and electric charge $Q$. The curves indicate that the presence of dark matter significantly modifies the thermal profile of the black hole, especially in the small-horizon regime. For each nonzero value of $\lambda$, the temperature starts from zero at a minimal radius, then increases until it reaches a maximum value, and from which the Hawking temperature decreases monotonically as $r_+$ becomes large. This behavior signals the existence of two branches: a small black hole branch, where the temperature increases with $r_+$, and a large black hole branch, where the temperature decreases with $r_+$.  
\\

The effect of increasing $\lambda$ is clearly to raise the temperature and shift the peak toward smaller horizon radii, which means that the dark matter environment enhances the thermal activity of the black hole. This enhancement is much more pronounced for small black holes, whereas for large $r_+$ all curves gradually approach one another, showing that the influence of dark matter becomes weaker in the large black hole phase. Therefore, Fig.~\ref{fig:trn} suggests that the dark matter parameter mainly affects the near-horizon thermodynamics of small black holes, while its effect on large black holes is subleading. 
\\

The electric charge $Q$ also plays an important role in the thermal behavior of the black hole. As $Q$ increases, the extremal horizon radius shifts to larger values, causing the physical temperature curves to start at larger $r_+$. At the same time, the maximum Hawking temperature decreases significantly, indicating that the electric charge suppresses the thermal activity of the black hole. This effect is particularly evident when comparing the panels corresponding to $Q=0.6$ and $Q=1.2$, where the temperature peak becomes lower and broader as the charge increases. However, variations in the electric charge do not affect the phase structure of the black hole. In all cases, two distinct phases are present: a small black hole phase and a large black hole phase. Physically, the electric charge counteracts the gravitational attraction and drives the system closer to the extremal configuration, thereby reducing the Hawking temperature. Consequently, dark matter and electric charge have opposite effects on the thermal properties of the black hole: increasing $\lambda$ enhances the temperature, whereas increasing $Q$ suppresses it.
\\

To study how dark matter affects the stability of RN black holes, we evaluate the expression of the heat capacity, which is given by  
\begin{equation}
	C= T_{RN}\, \frac{\partial S}{\partial T_{RN}}= \frac{2\pi r_+^2 \left(Q^2 - r_+ (r_+ + \lambda)\right)}{3Q^2 - r_+ (r_+ + 2\lambda)}.
\end{equation}

\begin{figure}[htp]
	\centering
	\includegraphics[width=0.45\linewidth]{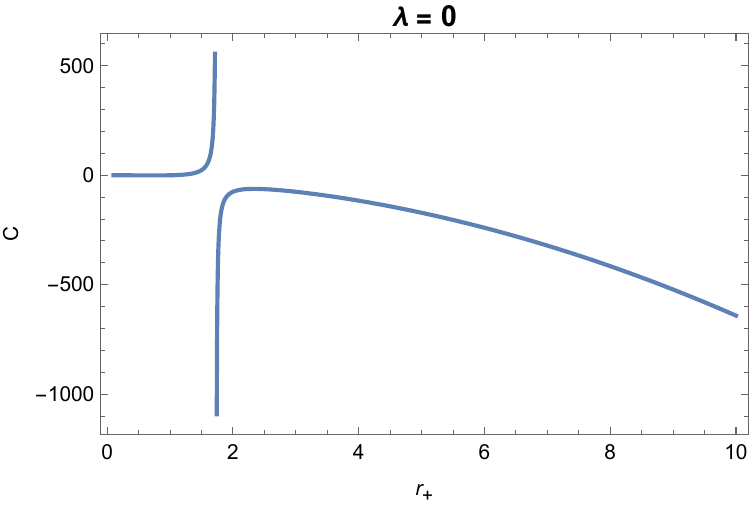}
	\includegraphics[width=0.45\linewidth]{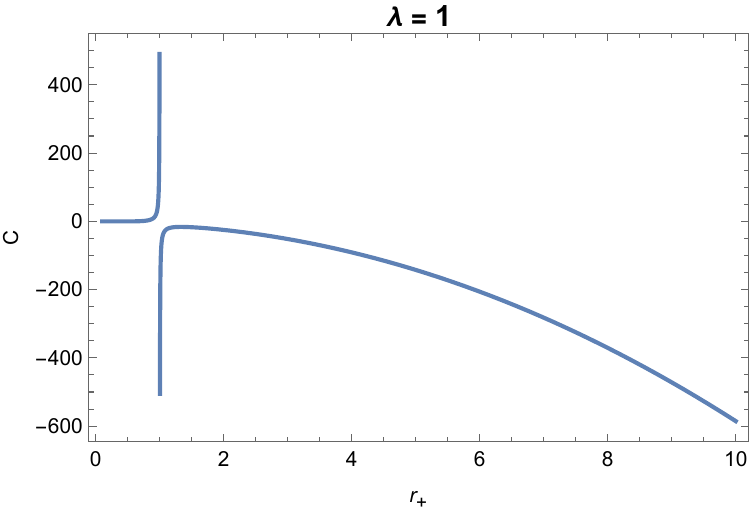}
	\includegraphics[width=0.45\linewidth]{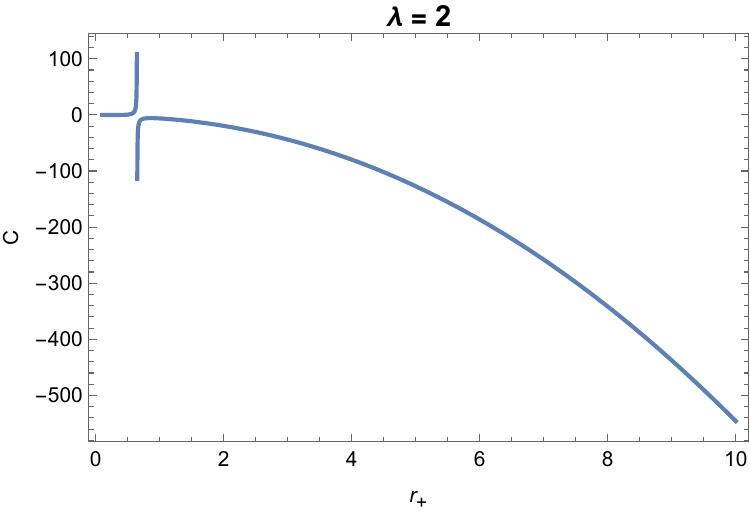}
	\includegraphics[width=0.45\linewidth]{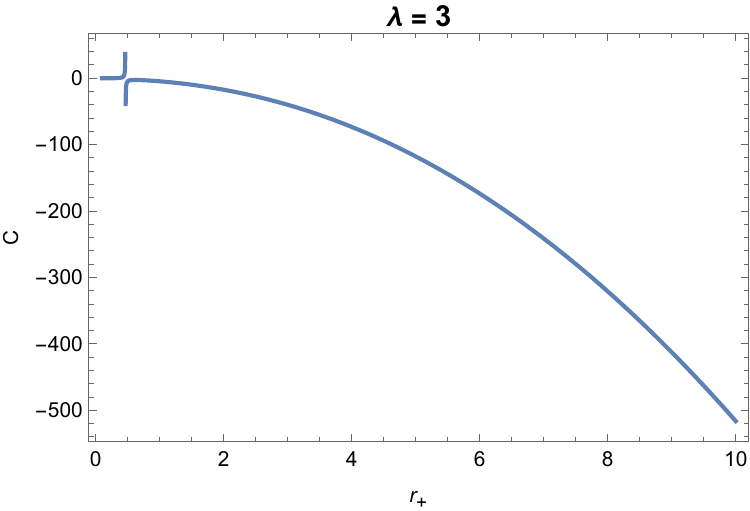}
	\caption{Heat capacity curves of RN black holes for different values of $\lambda$ and $Q=1$.}
	\label{fig:crn}
\end{figure}
Fig.~\ref{fig:crn} represents the heat capacity of RN black holes surrounded by dark matter for different values of the dark matter parameter $\lambda$ and for a fixed electric charge $Q=1$. The electric charge is fixed because variations in $Q$ do not modify the phase structure of the black hole. We observe that, in all cases, small black holes have a positive heat capacity, indicating that they are thermodynamically stable. In contrast, large black holes always have a negative heat capacity and are therefore thermodynamically unstable. The heat capacity diverges at the transition event horizon, signaling a phase transition between the small and large black hole phases. Furthermore, increasing the dark matter parameter $\lambda$ shifts the phase transition point but does not alter the overall phase structure or stability properties of the black hole. Therefore, the presence of dark matter does not change the qualitative thermodynamic behavior of RN black holes, which always exhibit two distinct phases: a stable small black hole phase and an unstable large black hole phase.

\section{Island Formula}
\label{sa}
In this section, we investigate the entanglement entropy of Hawking radiation emitted from Schwarzschild and RN black holes surrounded by perfect fluid dark matte via the island formula, to resolve the information loss paradox, and to study how dark matter around black holes affects information recovery.
\\

In the island formula, we first proceed by calculating the entanglement entropy without considering a region inside the black hole called the island. This calculation is performed before a characteristic time, called the Page time. In the second step, after the Page time, we calculate this entropy by including the contribution of the island. In this case, the entanglement entropy of Hawking radiation is given by \cite{18aa}
\begin{equation}
	\label{ife}
	S(R) = \min \left\{ \operatorname*{ext} \left[ \frac{\mathrm{Area}(\partial I)}{4} + S_{\mathrm{Bulk}}(R \cup I) \right] \right\},
\end{equation}
here, $R$ represents the region of Hawking radiation outside the black hole, $I$ is the island, and $\partial I$ represents the boundary of the island. Furthermore, the terms ``ext'' and ``min'' mean that the generalized entropy must be extremized with respect to the position of the island boundary, $\partial I$, and that, among all possible extremal configurations, the physical entropy corresponds to the minimum value, respectively.
\\

To study information recovery through the island formula for black holes surrounded by perfect fluid dark matte, we use a regular coordinate system across the event horizon, namely the Kruskal coordinates. These coordinates are given by \cite{Lin:2024gip}
\begin{equation}
	\label{ife2}
	U = -e^{-\kappa (t - r_*)}, \qquad V = e^{\kappa (t + r_*)},
\end{equation}
where $\kappa$ is the surface gravity at the event horizon, and $r_*$ represents the tortoise coordinate, which is expressed as
\begin{equation}
	r_* = \int \frac{1}{f(r)} \, dr.
\end{equation}
Besides using these coordinates, by considering the large-distance limit and employing the $s$-wave approximation, we neglect the angular part of the metric, and the metric becomes two-dimensional, as
\begin{equation}
	ds^2 = W(r)^2 dU\, dV,
\end{equation}
where $W(r)$ represents a conformal factor, which can be expressed as follows
\begin{equation}
	W(r)^2 = \frac{f(r)}{\kappa^2 e^{2\kappa r_*}}.
\end{equation}
\subsection{Schwarzschild Black Holes}
To investigate the entanglement entropy of Hawking radiation for Schwarzschild black holes surrounded by perfect fluid dark matter via the island formula, we proceed as mentioned above.

\subsubsection{Without Island}
\label{Ss1}
We start our analysis by calculating the entanglement entropy before the Page time, without the consideration of the island.

\begin{figure}[htp]
	\centering
	\includegraphics[width=0.5\linewidth]{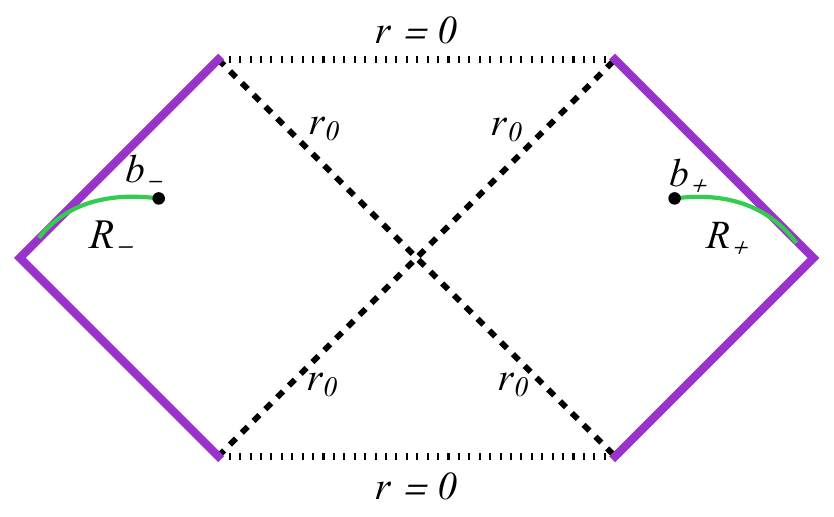}
	\caption{Penrose diagram of a Schwarzschild black hole surrounded by perfect fluid dark matte without an island.}
	\label{si}
\end{figure}

Fig. \ref{si} represents the Penrose diagram of a Schwarzschild black hole surrounded by dark matter without an island. In this figure, we present the Hawking radiation by the region $R$ in green color, where this region has two parts, $R_+$ and $R_-$, located in the right and left wedges, respectively. Furthermore, $b_\pm$ represent the boundary surfaces of these regions, and correspond to the following coordinates: $(t_b, b)$ for $b_+$ and $(-t_b + i \beta/2,b )$ for $b_-$, with $\beta$ denotes the inverse of the Hawking temperature.
\\

Initially, $t=0$, we assume that the global system is in a pure state. For this reason, the von Neumann entropy of a subsystem is equal to the entropy of its complementary region. This property is called the complementarity principle of the von Neumann entropy \cite{Lin:2024gip}.
\\

The region $R$ of Hawking radiation is defined as follows
\begin{equation}
	R=  ] - \infty, b_- [\quad \text{U}\quad [  b_+, +\infty [.
\end{equation}

By using the complementarity principle, we can calculate the von Neumann entropy of the region $R$, where its complementary region is $[b_-, b_+]$. This region is described by the conformal field theory. Therefore, the entropy of $[b_-, b_+]$ is equivalent to the entanglement entropy of Hawking radiation. We express this entropy as follows
\begin{equation}
	\label{SRE}
	S(R)= \frac{C}{3} \log[l(b_+,b_-)]
\end{equation}
where $C$ denotes the conformal field theory central charge and $l(b_+,b_-)$ represents the geodesic distance, which is given by \cite{19}
\begin{equation}
	\label{SRD}
	l^2(b_{+}, b_{-}) = W^2(b)[U(b_{+}) - U(b_{-})][V(b_{-}) - V(b_{+})].
\end{equation}
The conformal factor can be written in the large-radial-distance limit as
\begin{equation}
	\label{SRS}
	W^2(b)= \frac{e^{-2 \kappa r_*}}{\kappa^2}.
\end{equation}

Through Eqs.~\eqref{SRE}-\eqref{SRS}, we find the entanglement entropy of Hawking radiation as 
\begin{equation}
	S(R)= \frac{C}{3} \log\left(\frac{2 \cosh \left(\kappa t_b \right) }{\kappa } \right),
\end{equation}

For late times, this entropy can be written as follows
\begin{equation}
	\label{SRF}
	S(R)= \frac{C}{3} \kappa t_b.
\end{equation}

From the final expression of the entanglement entropy of Hawking radiation for Schwarzschild black holes surrounded by perfect fluid dark matte, we observe that this entropy diverges at late times, which violates the Page curve. This means the loss of information in this phase.
\subsubsection{With Island}
\label{Ss2}
To resolve the issue of the divergence of the entropy, and the loss of information, we include the contribution of the island in the entanglement entropy of Hawking radiation after the Page time.

\begin{figure}[htp]
	\centering
	\includegraphics[width=0.5\linewidth]{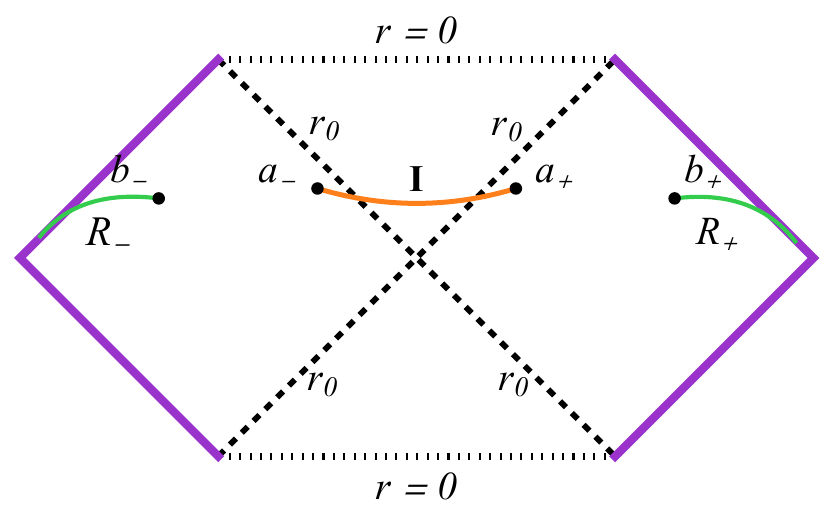}
	\caption{Penrose diagram of a Schwarzschild black hole surrounded by perfect fluid dark matte with an island.}
	\label{swi}
\end{figure}

Fig. \ref{swi} represents the Penrose diagram of a Schwarzschild black hole surrounded by perfect fluid dark matte, with the consideration of the island represented by the region $I$ in orange color, and $a_\pm$ represent the boundaries of the island, where $a_+$ corresponds to the coordinates $(t_a,a)$ and $a_-$ corresponds to $(-t_a+ i \beta/2,a)$.
\\

From the generalized entropy and the conditions of extremality and minimality, we can calculate the entanglement entropy with the consideration of the island contribution. The generalized entropy can be expressed as follows \cite{Lin:2024gip}
\begin{equation}
	\label{SSRT}
	S_{\mathrm{gen}}(R)
	=
	2\pi a^2
	+
	\frac{C}{3}
	\log
	\frac{
		l(a_{+},a_{-})
		l(b_{+},b_{-})
		l(a_{+},b_{+})
		l(a_{-},b_{-})
	}{
		l(a_{+},b_{-})
		l(a_{-},b_{+})
	}.
\end{equation}

This expression can be simplified by considering that the distance between the left and right wedges is very large, so we find that \cite{19}
\begin{equation}
	l(a_{+},a_{-}) \simeq l(b_{+},b_{-})
	\simeq l(a_{+},b_{-})
	\simeq l(a_{-},b_{+})
	\gg l(a_{+},b_{+})
	\simeq l(a_{-},b_{-}) .
\end{equation}

By using this approximation, the generalized entropy becomes as follows
\begin{equation}
	\label{soe}
	S_{\mathrm{gen}}(R)
	= 2\pi a^2
	+ \frac{C}{3}
	\log \Bigl[
	l(a_{+},b_{+})\,l(a_{-},b_{-})
	\Bigr].
\end{equation}

Through Eq. \eqref{soe}, we find the following expression for the generalized entropy
\begin{equation}
	S_{\mathrm{gen}}(R)=2\pi a^2+\frac{C}{3}\log\left\{\frac{2\sqrt{f(a)}}{\kappa^2}\left[\cosh\bigl(\kappa(r_*(a)-r_*(b))\bigr)-\cosh\bigl(\kappa(t_a-t_b)\bigr)\right]\right\}.
\end{equation}
\\

Through the inclusion of the minimal and extremal conditions, we can find the final expression of the entanglement entropy of Hawking radiation with an island. To find the coordinates, we express the extremal condition as $\frac{\partial S_{\mathrm{gen}}}{\partial t_a} =
\frac{\partial S_{\mathrm{gen}}}{\partial a} = 0$. By solving $\frac{\partial S_{\mathrm{gen}}}{\partial t_a}=0$,  the temporal coordinates of the island, where we obtain $t_a=t_b$. For the spatial coordinate of the island, we have that the island lies close to the black hole event horizon, so we can write the spatial coordinate, $a$, in terms of the event horizon, as follows
\begin{equation}
	a= r_0 + \epsilon^2 r_0
\end{equation}
where $\epsilon \ll 1$. From this approximation, we can also write the metric function at the island spatial coordinate as follows
\begin{equation}
	\label{4X}
f(a)\approx f'(r_0)(a-r_0)= 2 \kappa r_0 \epsilon^2,
\end{equation}
and the tortoise coordinate is expressed as
\begin{equation}
	\label{5X}
	r_*(a)= \frac{1}{\kappa} \log\left( \epsilon\right),
\end{equation}

By using Eqs. \eqref{4X} and \eqref{5X}, and the spatial extremal condition, $\frac{\partial S_{\mathrm{gen}}}{\partial a}=0$, we find the expression of the spatial coordinate of the island as follows
\begin{equation}
	a= r_0+ \left(\frac{C e^{-\kappa r_*(b)}}{12 \pi r_0^2} \right)^2 r_0.
\end{equation}
where
\begin{equation}
	\epsilon= \frac{C e^{-\kappa r_*(b)}}{12 \pi r_0^2}.
\end{equation}

For the condition of minimality, we have only one value of the island coordinates, so there is only one value of the entanglement entropy of Hawking radiation.
\\

Finally, by using this result, we can express the entanglement entropy of Hawking radiation with an island for a Schwarzschild black hole surrounded by perfect fluid dark matte as follows
\begin{equation}
	S(R)
	= 2S + 4\epsilon^{2}S
	+ \frac{C}{6}
	\log\!\left(
	\frac{
		2r_0 \left(
		e^{2k r_{*}(b)}
		- 4\epsilon\, e^{k r_{*}(b)}
		+ 6\epsilon^{2}
		\right)
	}{
		\kappa^{3}
	}
	\right).
\end{equation}

We can write the final expression of the entropy in the following approximation
\begin{equation}
	S(R)=2 S_{BH}.
\end{equation}

We observe that the resulting entropy is independent of time, where it is fixed at two times the Bekenstein-Hawking entropy. This means that the entanglement entropy after the Page time is constant due to the contribution of the island. In general, we have two phases of the entanglement entropy. The first is before the Page time, where the entropy increases with time. The second phase is after the Page time, where the entanglement entropy is independent of time and fixed at $2 S_{BH}$. As a result, the general evolution of the entanglement entropy of Hawking radiation follows the Page curve. Therefore, before the Page time, a loss of information occurs, and after the Page time, a recovery of information occurs. We conclude that the island formula is valid for information recovery for Schwarzschild black holes surrounded by perfect fluid dark matte.

Also, we have that the information recovery is related to the Page time. Through the equality between the entanglement entropy without and with an island at the Page time, we find the expression of the Page time as follows
\begin{equation}
	\label{PTS}
	t_P= \frac{3 r_0^2}{C T},
\end{equation}

We observe that the Page time is directly related to the Hawking temperature and the event horizon of the black hole. This means that there is a correspondence between black hole thermodynamics and information recovery, which we will discuss in more detail in the next section.  
\subsection{Reissner-Nordström Black Holes}

We now extend our analysis to Reissner-Nordström (RN) black holes surrounded by perfect fluid dark matte. Using the island formula, we investigate the entanglement entropy of Hawking radiation before and after the Page time, following the same methodology for Schwarzschild black holes.
\\

Unlike the Schwarzschild case, RN black holes possess an extremal configuration characterized by the coincidence of the inner and outer horizons. At the extremal limit, corresponding to $r_+=r_{\mathrm{ext}}$, the Hawking temperature vanishes and the black hole ceases to radiate. Consequently, Hawking radiation is absent, the entanglement entropy does not evolve, and the information loss paradox does not arise. For this reason, our analysis is restricted to non-extremal RN black holes satisfying $r_+>r_{\mathrm{ext}}$, for which Hawking radiation is emitted and information recovery can be meaningfully investigated.
\

We begin by evaluating the entanglement entropy of Hawking radiation in the absence of an island.

\begin{figure}[htp]
	\centering
	\includegraphics[width=0.5\linewidth]{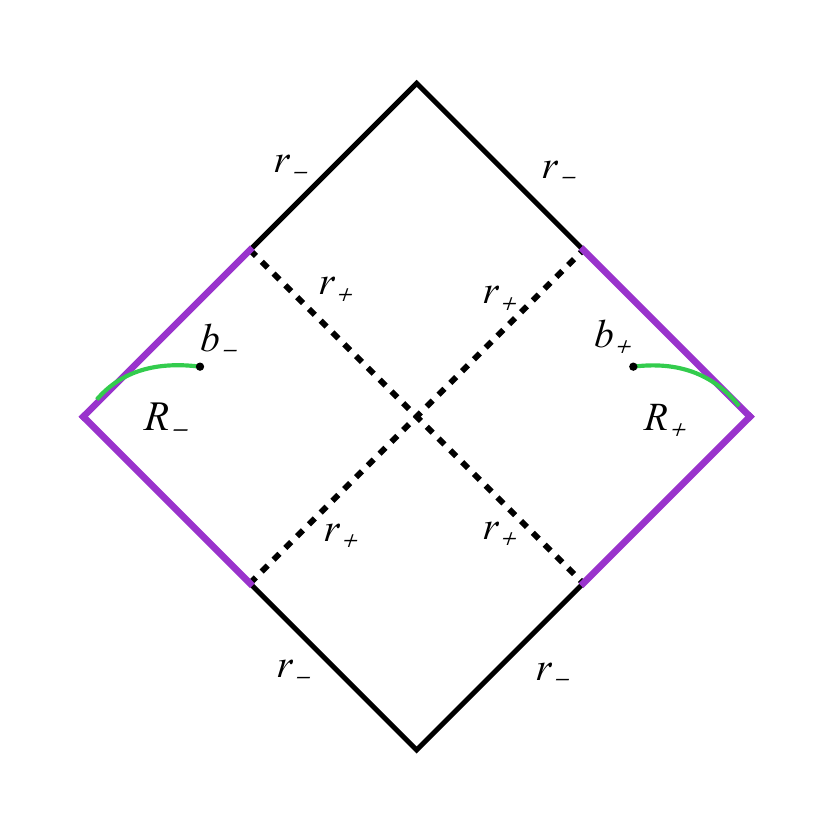}
	\caption{Penrose diagram of a RN black hole surrounded by perfect fluid dark matte without an island.}
	\label{RNWI}
\end{figure}

Fig.~\ref{RNWI} depicts the Penrose diagram of a RN black hole surrounded by perfect fluid dark matte. Here, $r_+$ and $r_-$ denote the event horizon and the inner (Cauchy) horizon, respectively. The green regions $R_\pm$ represent the Hawking radiation, while $b_\pm$ denote the corresponding boundary surfaces. Their coordinates are given by $(t_b,b)$ for $b_+$ and $(-t_b+i\beta/2,b)$ for $b_-$. Following the same procedure presented in Subsubsection~\ref{Ss1}, we obtain

\begin{equation}
	S(R)=\frac{C}{3}\log\left(\frac{2\cosh(\kappa t_b)}{\kappa}\right).
\end{equation}

In the late time regime, this expression reduces to

\begin{equation}
	S(R)=\frac{C}{3}\kappa t_b.
\end{equation}

Therefore, the entanglement entropy grows linearly with time and diverges at late times. As in the Schwarzschild case, this behavior is inconsistent with the Page curve and signals an apparent loss of information within the semiclassical description.
\

To resolve this issue, we  include the contribution of the island and evaluate the entanglement entropy after the Page time.

\begin{figure}[htp]
	\centering
	\includegraphics[width=0.5\linewidth]{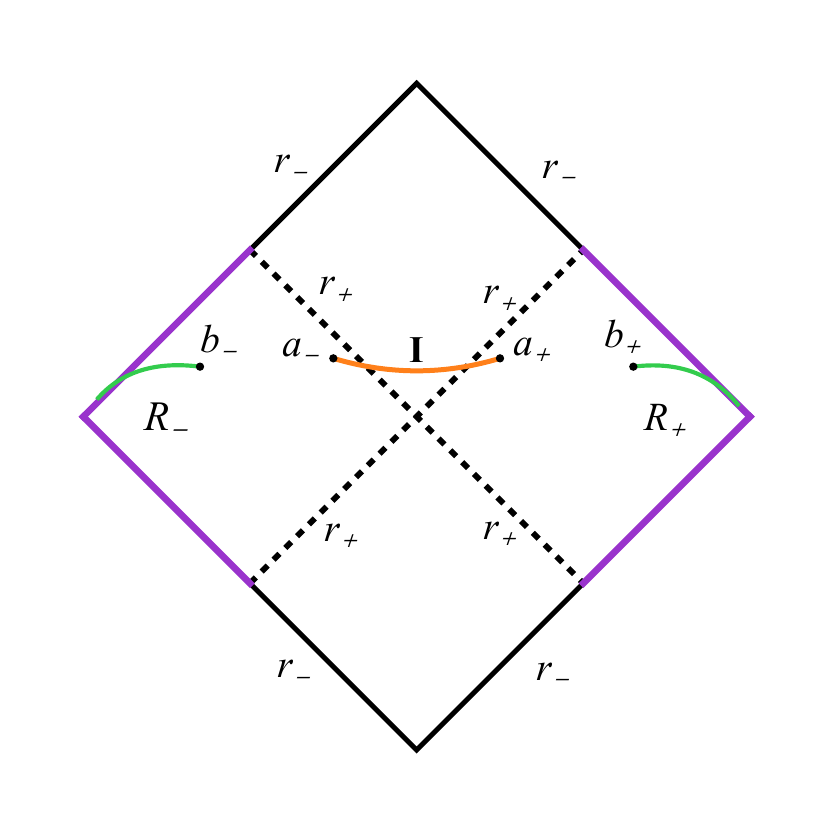}
	\caption{Penrose diagram of a RN black hole surrounded by perfect fluid dark matte with an island.}
	\label{RNI}
\end{figure}

Fig.~\ref{RNI} illustrates the corresponding Penrose diagram when an island is taken into account. The orange region denotes the island $I$, while $a_\pm$ represent its boundaries. The coordinates of these boundaries are $(t_a,a)$ for $a_+$ and $(-t_a+i\beta/2,a)$ for $a_-$. Applying the same analysis as in Subsubsection~\ref{Ss2}, we obtain

\begin{equation}
	S(R)=2S_{BH}.
\end{equation}

We therefore find that the entanglement entropy increases during the early stage of evaporation and becomes constant after the Page time, where it is fixed at twice the Bekenstein-Hawking entropy. Consequently, the complete evolution of the entropy follows the Page curve. The initial phase is characterized by an apparent loss of information, whereas the late-time phase corresponds to information recovery. The island contribution thus restores the unitary evolution of the evaporation process and provides a resolution of the information loss paradox for RN black holes surrounded by perfect fluid dark matte.

The Page time is obtained by equating the entanglement entropy with and without an island at the transition point. This yields

\begin{equation}
	\label{PTSS}
	t_P=\frac{3r_+^2}{CT}.
\end{equation}

\section{Information-Thermodynamics Correspondence}
\label{saa}
In this section, we investigate the signatures of dark matter in information recovery. Furthermore, we explore the relationship between the thermodynamic properties of black holes and the information recovery process through the Page time.

In the previous sections, we demonstrated that the Page time for Schwarzschild and RN black holes surrounded by perfect fluid dark matte is directly determined by the event horizon radius and the Hawking temperature (see Eqs.~\eqref{PTS} and \eqref{PTSS}). This result indicates that the information recovery process is intrinsically connected to the thermodynamic properties of black holes. Consequently, the Page time encodes valuable information regarding both the surrounding dark matter distribution and the thermodynamic state of the black hole.

\subsection{Schwarzschild Black Holes}

We now investigate the effects of dark matter and black hole thermodynamics on information recovery for Schwarzschild black holes surrounded by perfect fluid dark matte. Combining Eqs.~\eqref{Ta} and \eqref{PTS}, the Page time can be expressed in terms of the black hole parameters as

\begin{equation}
	t_P = \frac{12 \pi r_0^4}{C(\lambda+r_0)}.
\end{equation}

This expression provides important insight into the connection between black hole evaporation and information recovery. The Page time corresponds to the moment at which the entanglement entropy of Hawking radiation reaches its maximum value and the island contribution becomes dominant. Physically, it marks the transition from the information-loss phase to the information-recovery phase. Therefore, the Page time plays a central role in the study of black hole information dynamics. In the following, we analyze the influence of the black hole size and the dark matter parameter on the behavior of the Page time.

\begin{figure}[htp]
	\centering
	\includegraphics[width=0.7\linewidth]{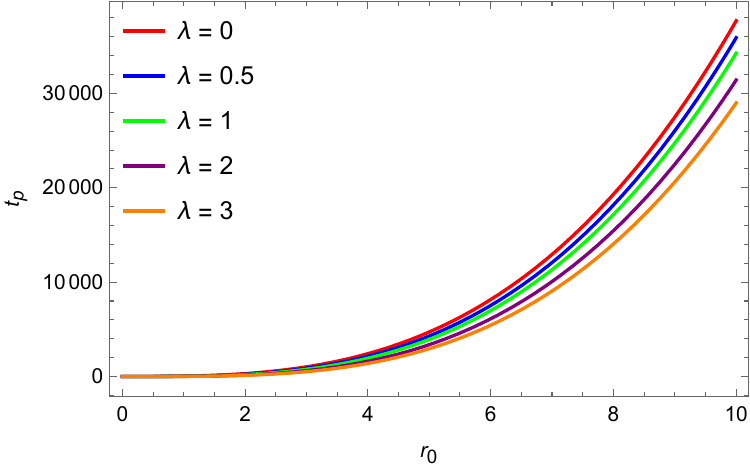}
	\caption{Page time as a function of the event horizon radius for Schwarzschild black holes surrounded by perfect fluid dark matte for different values of the dark matter parameter $\lambda$.}
	\label{tpsh}
\end{figure}

Fig.~\ref{tpsh} illustrates the dependence of the Page time on the event horizon radius for several values of the dark matter parameter $\lambda$. We observe that the Page time increases monotonically with the event horizon radius. Consequently, larger black holes require a significantly longer time to reach the onset of information recovery than smaller black holes. This behavior reflects the fact that the information recovery process is strongly influenced by the size of the black hole, with small black holes recovering information more rapidly than their larger counterparts.

The figure also reveals the influence of the dark matter parameter on the Page time. For small black holes, the effect of $\lambda$ is relatively weak. However, as the event horizon radius increases, the impact of dark matter becomes increasingly significant. In particular, increasing the dark matter parameter leads to a systematic decrease in the Page time. This behavior indicates that the presence of perfect fluid dark matte modifies the dynamics of information recovery by reducing the Page time, thereby accelerating the onset of the information recovery phase.

From a physical perspective, the reduction of the Page time can be understood through the influence of perfect fluid dark matte on the Hawking temperature. As shown in the thermodynamic analysis, the dark matter parameter enhances the Hawking temperature of Schwarzschild black holes. Since the Page time is inversely proportional to the Hawking temperature, an increase in temperature naturally results in a shorter Page time. Consequently, black holes embedded in denser dark matter environments begin to recover information earlier than isolated black holes.

\subsection{Reissner-Nordström Black Holes}

We now extend our investigation of the information-thermodynamics correspondence and the effects of dark matter to RN black holes surrounded by perfect fluid dark matte. In contrast to Schwarzschild black holes, RN black holes possess a richer thermodynamic structure due to the presence of the electric charge, including an extremal limit and distinct thermodynamic phases. Consequently, they provide an ideal framework for exploring the interplay between thermodynamic phase behavior and information recovery.

To analyze this correspondence, we first derive the analytical expression for the Page time. Combining Eqs.~\eqref{TRNz} and \eqref{PTSS}, we obtain

\begin{equation}
	t_P= \frac{48 \pi^2 r_+^5}{r_+\left(r_+ + \lambda\right)-Q^2}.
\end{equation}

\begin{figure}[htp]
	\centering
	\includegraphics[width=0.7\linewidth]{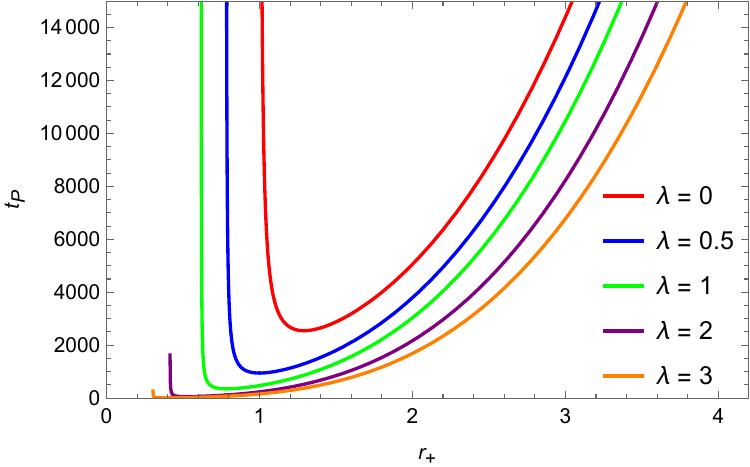}
	\caption{Page time as a function of the event horizon radius for RN black holes surrounded by perfect fluid dark matte for different values of the dark matter parameter $\lambda$ and fixed electric charge $Q=1$.}
	\label{tash}
\end{figure}

Fig.~\ref{tash} illustrates the Page time as a function of the event horizon radius for RN black holes surrounded by perfect fluid dark matte with fixed electric charge and different values of the dark matter parameter. We observe that the Page time exhibits two distinct branches, corrpondandce small and large black holes.

For black holes close to the extremal configuration, small black holes, the Page time decreases rapidly as the event horizon radius increases. After reaching a minimum value, the Page time begins to increase with the horizon radius, corresponding to the large black hole branch. Therefore, the Page time displays a non-monotonic dependence on the event horizon radius, reflecting the nontrivial thermodynamic behavior of RN black holes.

The effect of the dark matter parameter is also evident in Fig.~\ref{tash}. Increasing the value of $\lambda$ systematically shifts the Page time toward smaller values. Consequently, black holes surrounded by a denser perfect fluid dark matte environment begin the information recovery process earlier than those with smaller values of the dark matter parameter. This behavior indicates that perfect fluid dark matte modifies the dynamics of information recovery by reducing the Page time, thereby accelerating the onset of the information recovery phase.

These results reveal a direct correspondence between black hole thermodynamics and information recovery. In particular, the Page time serves as a bridge connecting the thermodynamic properties of the black hole to the dynamics of information retrieval, while simultaneously providing a potential signature of the surrounding dark matter environment. 
 
\section{Conclusion and Discussion}
\label{sec5}
In this paper, we studied the thermodynamic properties and information recovery of Schwarzschild and Reissner–Nordström black holes surrounded by perfect fluid dark matter. By incorporating the dark matter contribution through a logarithmic correction in the metric function, we analyzed how the surrounding dark matter environment affects black hole thermodynamics, evaporation, and the recovery of information through the island prescription.
\\

For Schwarzschild black holes, we found that the presence of dark matter increases the Hawking temperature, with a more pronounced effect for small black holes. Despite this modification, the entropy continues to satisfy the Bekenstein-Hawking area law, and the thermodynamic structure remains unchanged, characterized by a single unstable phase with negative heat capacity. Furthermore, by analyzing the evaporation process, we showed that dark matter enhances the radiation rate, leading to a reduction in the black hole lifetime. This result highlights the significant role of the environment in black hole evolution, especially at small scales.
\\

For Reissner-Nordström black holes, we observed similar effects of dark matter on the Hawking temperature, where it contributes positively and enhances thermal activity. The entropy again remains unchanged, preserving the area law. The phase structure exhibits a transition between small and large black holes, governed by the electric charge and the dark matter parameter. However, we found that dark matter does not introduce new critical behavior or second-order phase transitions. The heat capacity analysis shows that small black holes are thermodynamically stable, while large black holes remain unstable, and this qualitative behavior is not altered by the presence of dark matter. Instead, dark matter primarily affects the near-horizon thermodynamics, modifying the temperature profile and shifting the location of the phase transition.
\\

We then studied the black hole information loss paradox using the island formula. In the absence of islands, the entanglement entropy of Hawking radiation grows linearly with time and diverges at late times, reproducing the  loss of information. However, once island contributions are included, the entropy saturates at late times and follows the Page curve for both Schwarzschild and RN black holes surrounded by perfect fluid dark matte. This result demonstrates that the island mechanism remains effective in restoring unitary evolution even in the presence of a dark matter environment. A central result of this work is the establishment of a direct correspondence between black hole thermodynamics and information recovery. We showed that the Page time is determined by thermodynamic quantities such as the Hawking temperature and the horizon radius. Since perfect fluid dark matte increases the Hawking temperature, it  reduces the Page time and accelerates the onset of information recovery. 
\\

We conclude that perfect fluid dark matter leaves clear imprints on black hole evaporation, thermodynamics, and information recovery. These results reinforce the idea that realistic black holes cannot be treated as isolated systems and highlight the importance of environmental effects in the study of black hole physics, quantum gravity, and the black hole information paradox.
\\

Several extensions of this work can be considered. It would be interesting to investigate rotating black holes surrounded by perfect fluid dark matter and analyze how angular momentum modifies the thermodynamic behavior and information recovery. Furthermore, understanding the coupling between black holes and dark matter may also shed light on the nature of dark matter itself.
\section*{Acknowledgments}

Y. Ladghami gratefully acknowledges the support from the "PhD-Associate Scholarship – PASS" grant provided by the National Center for Scientific and Technical Research in Morocco, under grant number 42 UMP2023.

\end{document}